\documentclass[]{article}
\usepackage[T1]{fontenc}

\usepackage{amsmath}
\usepackage{amssymb}
\usepackage{latexsym,cite}
\usepackage{graphicx}
\usepackage{svg}
\usepackage{epstopdf}
\usepackage{multirow}

\usepackage[cmintegrals]{newtxmath}

\usepackage{balance}

\usepackage[caption=false,font=normalsize,labelfont=sf,textfont=sf]{subfig}

\usepackage{graphicx}
\usepackage{epstopdf}
\usepackage{cite}
\usepackage{multirow}
\usepackage{cleveref}
\usepackage{balance}
\usepackage{footnote}
\makesavenoteenv{tabular}
\usepackage{array}
\usepackage{dblfloatfix}
\usepackage{color}

\usepackage{graphicx}
\usepackage{epstopdf}
\usepackage{cite}
\usepackage{multirow}
\usepackage{cleveref}
\usepackage{balance}
\usepackage{footnote}
\makesavenoteenv{tabular}
\usepackage{array}
\usepackage{dblfloatfix}
\usepackage{color}

\usepackage{blindtext}
\date{}
\usepackage[left=2.30cm, right=2.30cm, top=2.30cm, bottom=2.30cm]{geometry}

\title{Two-Parametric Nyquist Pulses with Better Performance\\Based on Inverse Hyperbolic Functions}

\author{
	Songbing~Liang, and Stylianos D. Assimonis
	\thanks{The authors are with the Institute of Electronics, Communications and Information Technology (ECIT), Queen's University Belfast, Belfast, BT3 9DT, U.K., (e-mail: \{sliang07,~s.assimonis\}@qub.ac.uk).}
}

\begin{document}
	
	\maketitle
	
	\begin{abstract}
		In this article, three new inter-symbol interference (ISI)-free pulses with enhanced performance compared to the state-of-the-art are proposed and studied in terms of frequency and time domain characteristics.
		They are based on inverse hyperbolic functions and on the concept of inner and outer functions, which was first introduced by the authors.
		New pulses are two-parametric, i.e., their design depends only on the roll-off factor and the timing jitter parameter, and they outperform most of the well-known pulses reported in the literature, since they present lower error probability, smaller maximum distortion and wider eye-diagram.
	\end{abstract}


	\section{Introduction}
	
	{T}{he} raised-cosine (RC) pulse, proposed by Nyquist almost a century ago \cite{1928Nyquist}, is the most commonly used pulse in digital communication systems in order to minimize the inter-symbol interference (ISI) effect, due to the delay distortion, which is inherent on these types of systems. A strong research activity in the field of pulse shaping has been going on since then and many authors have proposed two-parametric (i.e., its shape depends only on the roll-off factor, $\alpha$, and the timing jitter parameter, $t/T$) \cite{2001Beaulieu,2011Assimonis,2004Beaulieu,2004Assalini,2008Assimonis,2009Assimonis,2010Assimonis} 
	and more than two-parameters Nyquist pulses
	\cite{alexandru2016isi,2005Chandan,mohri2009isi,2011Alexandru}, which outperform the RC pulse. Regarding the latter, the extra design parameters (i.e., except of $\alpha$, $t/T$), are obtained through various optimization techniques in order  ISI to be reduced for each roll-of-factor and timing jitter combination. 
	However, in practical digital transmission systems, although $\alpha$ can be predefined, timing jitter is a random variable and should be predicted. 
	Hence, in this work we focus on the design of new two-parametric pulses, which present better performance for every combination of $\alpha$ and $t/T$, consequently, we will not include in our performance analysis Nyquist pulses with more than two parameters, because such a  comparison would be unfair.
	
	In pulse shaping, inverse hyperbolic functions  were first utilized by Assalini and Tonello \cite{2004Assalini}, who namely proposed the \textit{flipped-hyperbolic secant} (fsech) and the \textit{flipped-inverse hyperbolic secant} (farcsech) pulses. 
	The concept of inner and outer functions was first introduced by Assimonis \textit{et al}. \cite{2008Assimonis,2011Assimonis}, by proposing two families of Nyquist ISI-free pulses.
	Specifically, in \cite{2011Assimonis} the ISI-free pulse cos[log], among others, was presented, which is based on the composition of the functions \textit{inverse cosine} (outer function) and  \textit{natural logarithm} (inner function), while in  \cite{2008Assimonis} the pulse acos[asinh] was presented, having the \textit{inverse cosine} and the \textit{inverse hyperbolic sine} function, as inner and outer function, respectively. 
	To the very best of our knowledge, these two ISI-free pulses present superior performance in terms of bit error rate (BER) among state-of-the-art of two-parametric Nyquist pulses. 
	
	In this work, we present three new two-parametric Nyquist ISI-free pulses, which are tested in terms of frequency and time domain characteristics and they present superior performance in terms of BER and eye-diagram compared to the state-of-the-art of two-parametric Nyquist pulses. 
	
	\section{New Two-Parametric Nyquist Pulses}
	
	\subsection{Frequency Domain Characteristics}
	
	Based on the inner and outer functions' concept \cite{2011Assimonis}, we proposed three new ISI-free pulses, whose  frequency response reads as, 
	\begin{equation}\label{puls}
		\begin{split}
			&S\left(f\right)= \\
			&\begin{cases}
				T, 																																	  &								         \left| f\right|  \le B\left(1-\alpha \right)  \\
				T\left\lbrace 1-\dfrac{1}{2\gamma}G\left(\dfrac{ \left| f\right|-B\left(1-\alpha \right)}{2\alpha B}\right)\right\rbrace,&B\left(1-\alpha \right) \le \left| f\right|  \le B  							  \\
				T\left\lbrace    \dfrac{1}{2\gamma}G\left(\dfrac{-\left|f\right|+B\left(1+\alpha \right)}{2\alpha B}\right)\right\rbrace,&B \le \left| f\right|  \le B\left(1+\alpha \right)  							  \\
				0, 																														         &  									 \left| f\right|  \ge B\left(1+\alpha \right) 	\\
			\end{cases}
		\end{split}
	\end{equation}
	where $ \alpha $ represents the roll-off factor ($ 0\le\alpha\le 1 $), i.e., the excess bandwidth beyond the Nyquist bandwidth and $ T $ denotes the symbol-period of the system, which defines the corresponding Nyquist frequency $ B=1/(2T) $. 
	Also, $ G\left( f\right) =g\left( h\left( f\right) \right)  $ is the composition function of the functions $ h(f) $ (inner) and $ g(f) $ (outer).
	In particular, the three new pulses are proposed according to 
	$ G(f)=\mathrm{acsch}(\log(f)) $, 
	$ G(f)=\mathrm{acoth}(\mathrm{acsch}(f)) $ and 
	$ G(f)=\mathrm{acsch}(\mathrm{asech}(f)) $,
	where, 
	$ \mathrm{acsch}$,  
	$ \mathrm{acoth} $ and
	$ \mathrm{asech} $
	represents 
	the \textit{inverse hyperbolic cosecant},
	the \textit{inverse hyperbolic cotangent} and
	the \textit{inverse hyperbolic secant} function,
	respectively,
	while 
	$ \log$
	is the \textit{natural logarithm} function.
	Continuity of \eqref{puls} implies that $ \gamma $ should meet the condition \cite{2011Assimonis}
	\begin{equation}
		\gamma=G\left({1}/{2}\right).
	\end{equation}
	Thus, for the first  (i.e., acsch[log]), 
	second (i.e., acoth[acsch]) 
	and 
	third (i.e., acsch[asech]) pulse $ \gamma $ constant is defined as
	%
	\begin{align}
		\gamma_{\mathrm{acsch[log]}} & =\mathrm{acsch}\left( \mathrm{log} 	\left( {1}/{2}\right) \right)\approx -1.16,  \\
		\gamma_{\mathrm{acoth[acsch]}} &=\mathrm{acoth}\left( \mathrm{acsch}	\left( {1}/{2}\right) \right)\approx  0.85, \\
		\gamma_{\mathrm{acsch[asech]}} &=\mathrm{acoth}\left( \mathrm{acsch}	\left( {1}/{2}\right) \right)\approx\ 0.7.
	\end{align}
	
\begin{figure}
		\centering
		\includegraphics[width=0.5\linewidth]{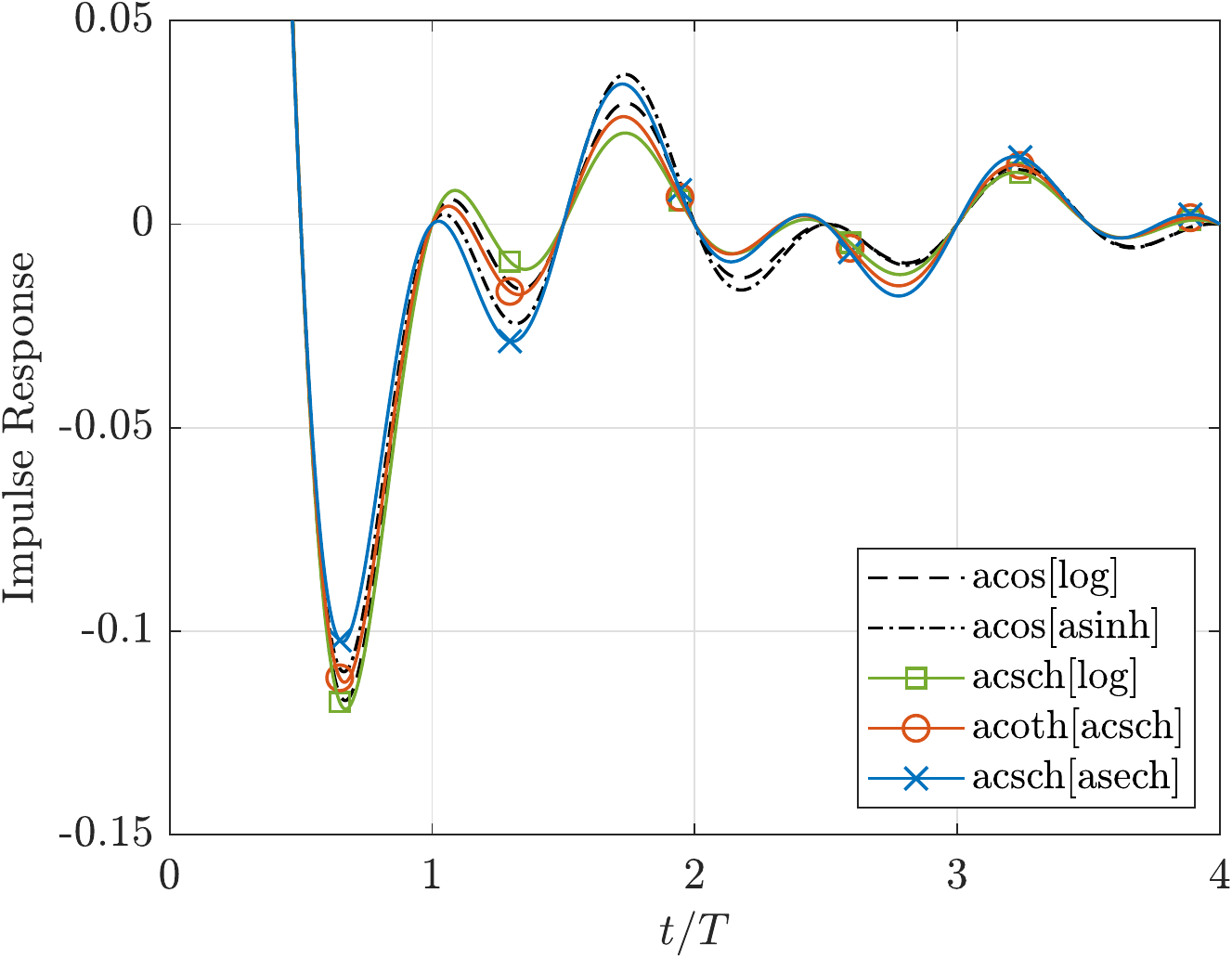}
		\caption{Impulse response of the reference and new pulses with roll-off factor $ \alpha=0.35 $.}
		\label{fig:impulse}
	\end{figure}

	The frequency responses of the three new pulses are illustrated in Fig. \ref{fig:impulse}. It is also depicted the acos[asinh] and acos[log] pulses for comparison purposes.
	It can be observed that all pulses' frequency response shows concave when $ B(1-\alpha)\le|f|\le B $, while convex in $ B\le|f|\le B(1+\alpha) $. Thus, they tend to transfer more power in higher-frequency region, which leads to lower first side-lobes in the time domain, as it will be shown, which in turn, leads to improved robustness against the time-jitter effects \cite{2011Assimonis}.

	\subsection{ Time-Domain Characteristics}
	
\begin{figure}
		\centering
		\includegraphics[width=0.5\linewidth]{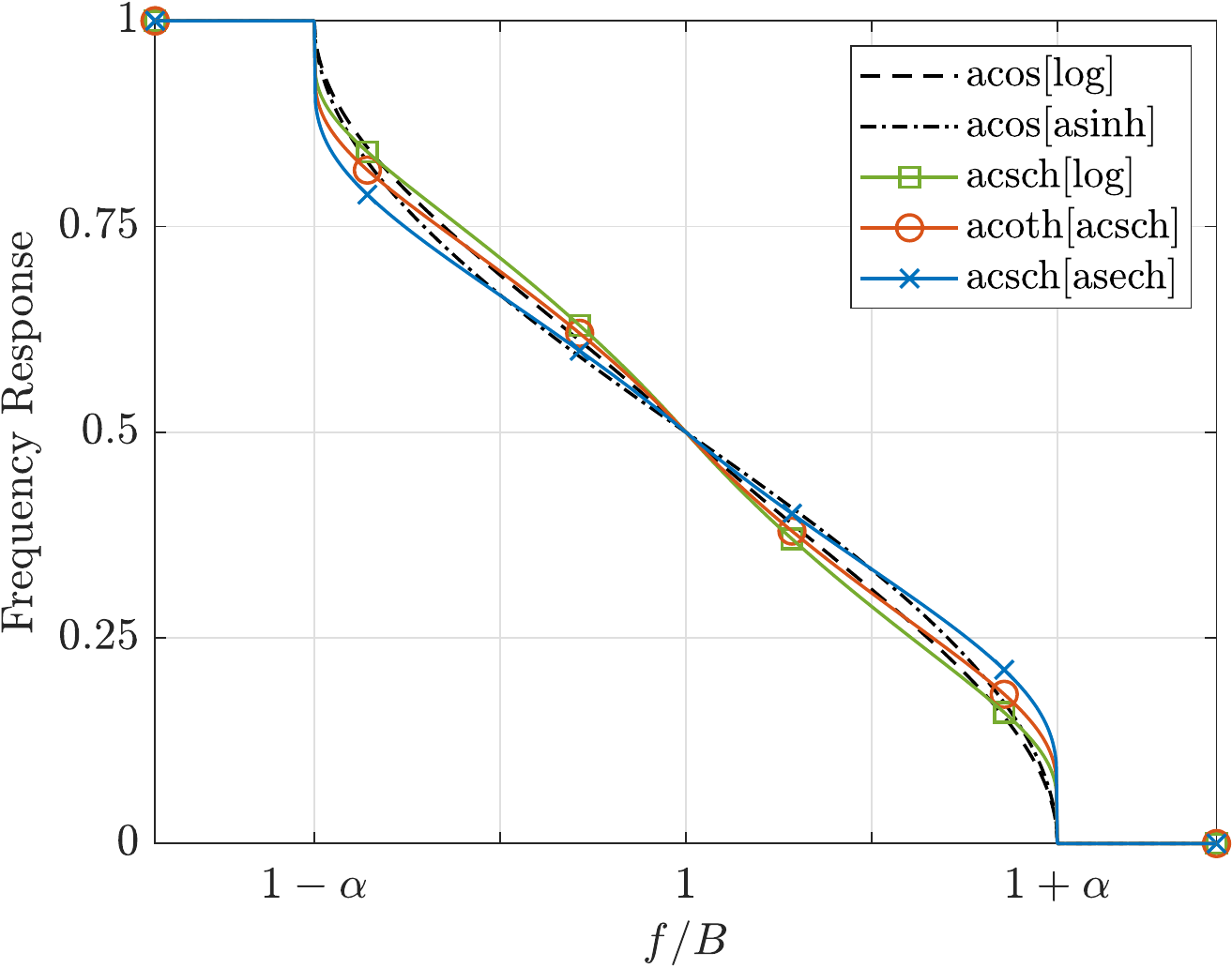}
		\caption{Frequency response of the reference and new pulses with roll-off factor $ \alpha=0.35 $.}
		\label{fig:freq}
	\end{figure}

	The impulse responses of the three new pulses are illustrated in Fig. \ref{fig:freq}, alongside with the acos[asinh] and acos[log] pulses for comparison sake. 
	It is evident that the acsch[asech] pulse have strong ability to suppress the first sidelobe. In general, the acsch[asech] is the best among all of five pulses considering the sidelobes suppression, while the acsch[log] pulse is not efficient in suppressing the sidelobes, although it presents the lower third and fourth sidelobe among these five pulses.
	It is noted that the new three pulses presented in this work and defined in the frequency domain in \eqref{fig:freq} do not have closed-form expression in time domain,  and hence, they have been numerically evaluated, similarly to the pulses presented in \cite{2004Assalini,2008Assimonis,2011Assimonis}.
	
	The decay rate can be estimated using the \textit{Theorem 1} introduced previously in \cite{2004Beaulieu}, among all these pulses. More specifically, we show that the following lemma is fulfilled.
	
	\textit{Lemma 1}: The decay rate of the pulse-family in \eqref{fig:freq} is $ 1/t^2 $ when $ \alpha\ne 0 $ and $ 1/|t| $ when $ \alpha=0 $.
	
	\textit{Proof}: For the shake of simplicity, we test in terms of time-domain properties only the pulse acsch[asech], while applying a similar procedure over the acsch[log] and acoth[acsch] will lead to exactly the same conclusion. 
	
	Assuming $ \alpha\ne{}0 $ the first (i.e., $ m=1 $) derivative of acsch[asech] is given by,
	\begin{equation}\label{dev_01}
		\begin{split}
			&S'\left(f\right)= \\
			&\begin{cases}
				0,
				&\left| f\right|  \le B\left(1-\alpha \right)\\
				T\dfrac{B \alpha}{
					\gamma c_1^2 
					\sqrt{
						c_1
						^2
						+1
					}
					\left(
					\mathrm{asinh}
					\left(
					c_1
					^2
					\right)
					-1
					\right)
				}
				,
				&B\left(1-\alpha \right) \le \left| f\right|  \le B\\
				T\dfrac{ B \alpha}{
					\gamma c_2^2 
					\sqrt{
						c_2
						^2
						+1
					}
					\left(
					\mathrm{asinh}
					\left(
					c_2
					^2
					\right)
					-1
					\right)
				},
				&B \le \left| f\right|  \le B\left(1+\alpha \right)  							  \\
				0
				,
				&\left| f\right|  \ge B\left(1+\alpha \right) 	\\
			\end{cases}
		\end{split}	
	\end{equation}
	where,
	%
	\begin{align}\label{dev_02}
		c_1 = \frac{2 B \alpha}{f - B\left(1-\alpha \right) }, \quad
		c_2 = \frac{2 B \alpha}{f - B\left(1+\alpha \right) }.
	\end{align}	

	Thus, based on \eqref{dev_01}, \eqref{dev_02}, we can readily observe that,
	\begin{equation}\label{dev_03}
		\lim\limits_{f\rightarrow B\left(1-\alpha\right) } S'\left(f\right) = 
		\lim\limits_{f\rightarrow B\left(1+\alpha\right) } S'\left(f\right) \rightarrow \infty
	\end{equation}
	and hence, discontinuity occurs at the transition points
	\begin{equation}\label{dev_04}
		f= B\left(1\mp\alpha\right).
	\end{equation}

	Assuming $ \alpha=0 $, the frequency response $ S(f) $ degenerates into a rectangular pulse with time-domain response the sinc function (i.e., $ \mathrm{sinc}(f)=\sin(t/T) $), which in turn, presents decay rate of $ 1/|t| $ \cite{2004Beaulieu,2011Assimonis}. The latter statement concludes the proof. \hfill $ \square $
	
	It is noted that despite the fact that the new pulses present lower decay rate than the well-known RC, which has a decay rate of $ 1/t^3 $, they present superior performance in terms of BER, as it will be shown next, since their first side-lobes amplitudes are lower than the RC: other studies \cite{2001Beaulieu,2004Assalini,2008Assimonis,2011Assimonis} came to the same conclusions.

	\begin{table}[t!]
		\renewcommand{\arraystretch}{1.25}
		\centering
		\caption{Eye width and max distortion of eye patterns}
		\label{table:eye-data}
		\begin{tabular}{c||cc}
			\hline
			pulse 		& eye-width & max distortion\\ \hline \hline
			acos[log]   & 0.780 	& 1.467 		\\
			acos[asinh] & 0.794 	& 1.475 		\\ 
			acsch[log]	& 0.812 	& 1.460 		\\
			acoth[acsch]& 0.802 	& 1.443 		\\
			acsch[asech]& 0.800 	& 1.440 		\\ \hline \hline
		\end{tabular}
	\end{table}
	
	\begin{table*}[t!]
		\renewcommand{\arraystretch}{1.25}
		\centering
		\caption{Bit error rate (BER) with $N=2^9$ Interfering Symbols and $ SNR=15 $ dB.}
		\label{table:BER-data}
		\begin{tabular}{l||l||cccc}
			\hline
			\multicolumn{1}{c||}{$\alpha$} & \multicolumn{1}{c||}{pulse} & $t/T= \pm 0.05$ & $t/T= \pm 0.1$ & $t/T=\pm 0.2$ & $t/T=\pm 0.3$ \\ \hline \hline
			\multirow{5}{*}{0.25}      
			& acos[log]                 & 5.3332e-8     & 1.0726e-6    & 2.7416e-4     & 1.2183e-2      \\
			& acos[asinh]               & 5.1488e-8     & 9.9816e-7    & 2.4946e-4     & 1.1349e-2      \\
			& acsch[log]                & 5.3700e-8     & 1.0861e-6    & 2.7826e-4     & 1.2345e-2      \\
			& acoth[acsch]              & 5.1677e-8     & 1.0003e-6    & 2.4874e-4     & 1.1393e-2      \\
			& acsch[asech]              & \textbf{4.9210e-8}     & \textbf{9.0533e-7}    & \textbf{2.1816e-4}     & \textbf{1.0306e-2}      \\ \hline 
			\multirow{5}{*}{0.35}      
			& acos[log]                 & 3.5470e-8     & 4.3365e-7    & 7.3486e-5     & 4.5509e-3      \\
			& acos[asinh]               & 3.4124e-8     & 4.0410e-7    & 6.7653e-5     & 4.2252e-3      \\
			& acsch[log]                & 3.5723e-8     & 4.3760e-7    & 7.3741e-5     & 4.5841e-3      \\
			& acoth[acsch]              & 3.4194e-8     & 4.0080e-7    & 6.5535e-5     & 4.1446e-3      \\
			& acsch[asech]              & \textbf{3.2414e-8}     & \textbf{3.6393e-7}    & \textbf{5.8702e-5}     & \textbf{3.7520e-3}      \\ \hline 
			\multirow{5}{*}{0.5}       
			& acos[log]                 & 2.1559e-8     & 1.4514e-7    & 1.4987e-5     & 1.2082e-3      \\
			& acos[asinh]               & 2.0758e-8     & 1.3617e-7    & 1.4609e-5     & 1.2202e-3      \\
			& acsch[log]                & 2.1693e-8     & 1.4525e-7    & 1.4541e-5     & 1.1712e-3      \\
			& acoth[acsch]              & 2.0726e-8     & 1.3274e-7    & 1.3347e-5     & \textbf{1.1118e-3}      \\
			& acsch[asech]              & \textbf{1.9693e-8}     & \textbf{1.2137e-7}    & \textbf{1.2947e-5}     & 1.1507e-3      \\ \hline \hline
		\end{tabular}
	\end{table*}

	\balance
	\subsection{Performance Evaluation}
	
\begin{figure}
		\centering
		\includegraphics[width=0.5\linewidth]{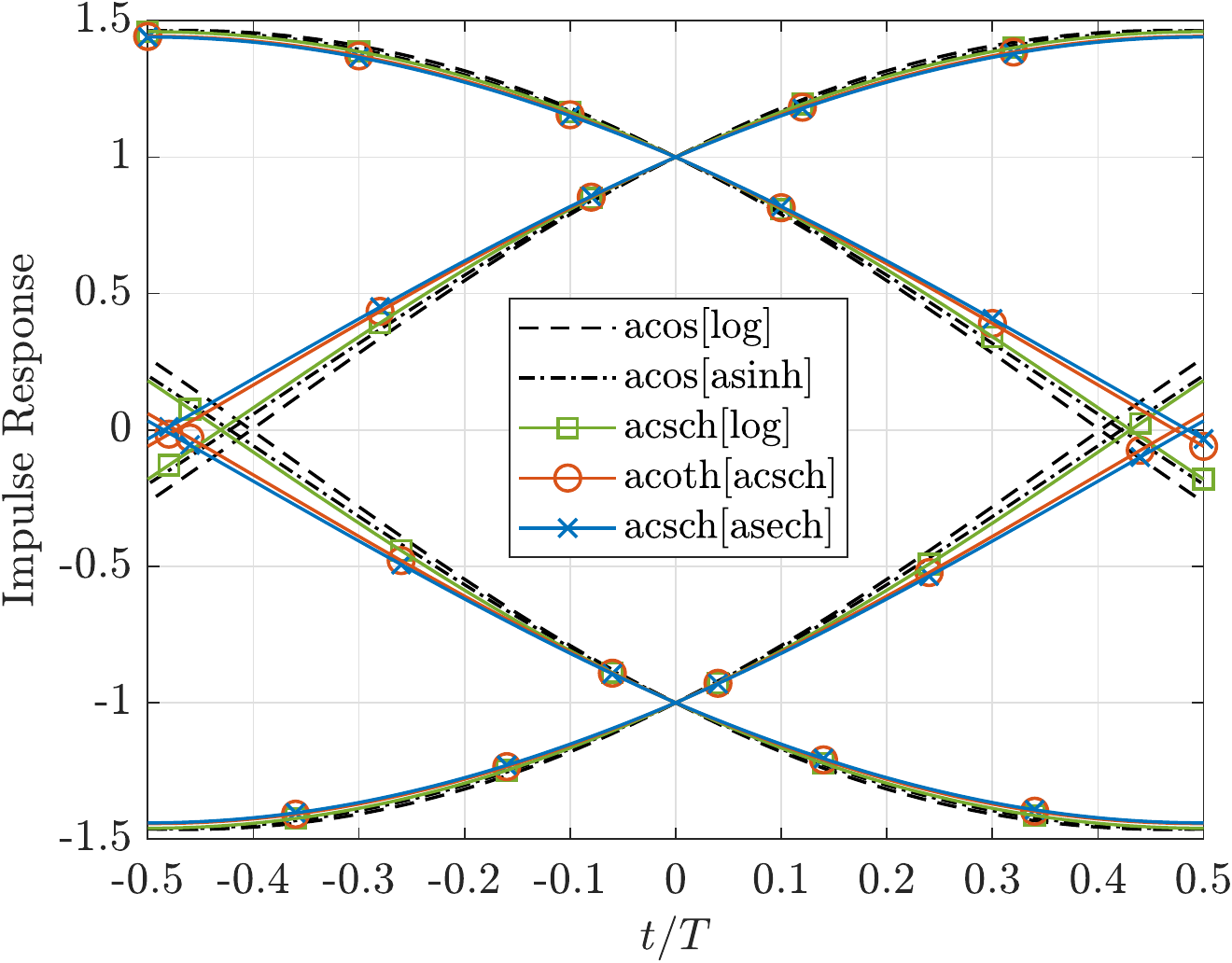}
		\caption{Inner and  outer boundaries of the eye-diagrams of the reference and the new pulses with roll-off factor $\alpha=0.35$.}
		\label{fig:eye}
	\end{figure}

	In this section, we evaluate the performance of the new three  Nyquist pulses in terms of eye-diagram and BER. In general, eye-diagram provides us with a great deal of visual information \cite{breed2005analyzing}, including the severity of ISI, sensitivity to timing jitter and the noise margin \cite{grami2015introduction}. 
	In our case, from the eye-diagram of the new pulses depicted in Fig. \ref{fig:eye}, we can obtain the their maximum distortion and the eye-width values, which are tabulated in Table \ref{table:eye-data}:
	it can be observed that the three new pulses present wider eye-width than the two reference pulses, which in turn leads to longer time interval whereinto the received signal can be sampled without the presence of ISI.
	Additionally, new pulses have better performance in terms of max distortion (Table \ref{table:eye-data}).

	A quantitative analysis can be specified by the BER estimation in the presence of time sampling errors. Please note that BER probability is the best measure of a digital system's performance, since includes the effects of noise, synchronization and distortion \cite{1291795}. 
	In this work BER was estimated according to \cite{120161}: we set parameters $ N_1=-100 $, $ N_2=100 $, signal-to-noise (SNR) ratio to $ 15 $ dB, number of the non-zero terms $ N_M=23 $ and $ N=2^9 $ interfering symbols. 
	The results are tabulated in Table \ref{table:BER-data}: this table gives the BERs of different pulses with the presence of timing offset (timing jitter $ t/T $ parameter) for varied roll-off factor $ \alpha $.  
	It can be easily noticed that, the new acsch[asech] pulse outperforms all other pulses in terms of BER (expect for the case of $ \alpha=0.5 $ and $ t/T = \pm 0.3$, where it finished second-best), and the BERs of acoth[acsch] and acos[asinh] are very close, taking second and third place at the most of the cases. On the other hand, reference pulses acos[log] and acsch[log]  rank last at the most of the cases.
	Please note that, the lowest BER is underlined (bold font) in Table \ref{table:BER-data} for each compination of $ \alpha $ and $ t/T $.

	Generally, it can be summarized that the three new Nyquist pulses always demonstrate an superior performance in terms of many aspects. In particular, based on the above analysis it can be noticed that the acsch[asech] is the best among all pulses we compared due to its lowest probability of getting distortion and the lower BER than other two-parametric Nyquist pulses. 
	Also, the dominant sidelobes are controlled better for the acsch[asech].
	
	\section{Conclusion}
	
	In this work, three new ISI-free Nyquist pulses were presented and studied in terms of frequency and time domain characteristics.  
	The design of these pulses is low-complexity, since it is based only on the roll-off factor and the timing jitter parameter (i.e., they are two-parametric pulses).
	They utilize the concept of the \textit{inner} and \textit{outer} functions and, also, they use inverse hyperbolic functions in their design.
	A fair comparison with the state-of-the-art of two-parametric pulses reveals their superiority in terms of performance.
	Specifically, to the best of our knowledge, the proposed pulses outperform all the currently published in the literature two-parametric Nyquist pulses it terms of BER and eye-diagram (i.e., maximum distortion, eye-width) and particularly, the acsch[asech] pulse is the best among all the considered pulses.

	%
	%
	%
	%
	%

	\bibliographystyle{IEEEtran}
	\bibliography{mybib.bib}

\end{document}